\documentclass[aps,epsfig,preprint]{revtex4}
\usepackage{color}
\usepackage{graphics}
\usepackage{epsfig}
\newcommand{\ts}{\textstyle}
\newcommand{\Pl}{\partial}

\newcommand{\bee}{\begin{equation}}
\newcommand{\ene}{\end{equation}}
\newcommand{\beea}{\begin{eqnarray}}
\newcommand{\enea}{\end{eqnarray}}

\newcommand{\fpar}[2]{\frac{{\ts \Pl \/ #1}}{{\ts \Pl \/ #2}}}

\begin{document}
\title{Propagation of slow electromagnetic disturbances in plasma}
\author{Sharad Kumar Yadav, Ratan Kumar Bera, Deepa Verma, Amita Das and Predhiman Kaw}
\affiliation{Institute for Plasma Research, Bhat , Gandhinagar - 382428, India }
\date{\today}
\begin{abstract}   
Electromagnetic (EM) waves/disturbances  are  typically the  best means to understand and analyze an ionized medium like plasma.
However, the propagation of electromagnetic waves with frequency lower than the plasma frequency is prohibited by the freely moving charges of the
plasma. In dense plasmas though the plasma frequency can be typically quite high,  EM sources at such higher frequency
are not easily available. It is, therefore, of interest to seek possibilities wherein
a low frequency (lower than the plasma frequency) EM disturbance  propagates inside a plasma. This is possible in the context of  magnetized plasmas.
However, in order to have a  magnetized plasma response  one requires a strong external magnetic field.
In this manuscript we demonstrate that the  nonlinearity of the plasma medium can also aid the propagation of a slow EM wave inside plasma.
Certain  interesting applications of
the propagation of such slow  electromagnetic pulse  through plasma is also discussed.
\end{abstract}
\pacs{} 
\maketitle 
\section{Introduction}
The  electromagnetic disturbances can get shielded by the moving free charges in the plasma \cite{dendy90}. The plasma contribution in the 
Maxwell's equations arises in the form of conduction current which tries to neutralize the displacement current and 
shields the EM wave  propagation through the  plasma. The conduction current in the plasma, however, has an upper limit of  $\mid J  \mid < enc$ 
(where $e$, $n$, and $c$ are electronic charge, electron number  density and the speed of light respectively) which 
cannot be exceeded under any circumstances. 
The  displacement current on the other hand would keep  increasing with the frequency of EM disturbance.  Thus a plasma system of a given density 
would be  able to shield the EM disturbance only up to a certain frequency.  This   exact condition translates to the 
well known fact  that plasma with a given density can shield 
radiation only up to  plasma frequency. Thus the EM radiation with frequencies less than the plasma frequency get reflected from the plasma. 
 This property 
 played a crucial role in  radio wave communication during  pre-satellite era. The radio wave reception was 
  possible even when transmitters and receivers where not along the  line of sight merely because the radio waves got 
  reflected from  the  ionospheric  plasma region. Nowadays, for  satellite which hover at higher altitudes than ionosphere, 
  microwaves are used for communication. Microwaves have a   higher frequency compared to the plasma frequency associated with the ionospheric plasma layer 
  and hence can propagate through the region  without reflection,  to reach the satellites \cite{dendy90}. 
  As one gets towards plasmas which are denser, seeking EM sources with frequencies higher than the plasma frequency for propagation 
becomes difficult. For instance, in the fast ignition concept \cite{tabak_1994} of laser fusion the ignitor laser pulse is unable to propagate beyond a density 
of $10^{21}/cc$ to deposit energy at the core with densities in the range of $\sim 10^{25}/cc $ for creating an ignition spark. It is , therefore, of interest 
to seek possibilities by which a low frequency EM disturbance can be made to propagate inside a dense plasma region. One clear solution is to 
have a magnetized plasma. For instance the Alfven wave \cite{freidberg87} and the Whistler waves \cite{kingsep,stenzel_prl08} with frequencies lower than the plasma frequency do  propagate inside 
the plasma. These EM waves, however,  require the presence of an externally applied  magnetic field in the medium. 
\par\vspace{\baselineskip}

 In this paper  we show that the nonlinearity of 
the plasma medium can  be exploited for the propagation of slow EM pulses in dense plasmas. In addition we also 
present certain interesting application of such EM disturbances. 
This includes a novel mechanism of  guiding, collimating and trapping of  the EM pulse by appropriate tailoring of the local plasma density profile. 
The  possibility of   splitting  a single EM pulse  and transporting them individually   to  distinct desired  locations in plasma 
has also been shown. These EM pulses are also  associated with 
certain electron current configurations in the plasma. Thus electron transportation also occurs along with the propagation of these  pulses. 
This mechanism  of electron transport can be viewed as an alternative to other recently proposed schemes which utilize specially  structured targets 
prepared of different materials having  varying  resistivity  transverse to the  propagation direction, for efficient transport \cite{robinson_pop,kar_prl}. 

\par\vspace{\baselineskip}

The manuscript has been organized as follows. Section II contains the description of such nonlinear EM disturbances which can be made to  propagate 
inside the dense plasma medium. Section III discusses the possibility of exploitation of these structures 
for novel applications  by suitably tailoring the plasma denisty profile. Section IV contains the summary and discussion.
\section{Propagation of slow  Electromagnetic  disturbances in high density plasmas}
It is well known that the electromagnetic wave with frequencies greater than the plasma frequency  only can propagate inside plasma if the plasma is unmagnetized \cite{dendy90}. 
In the presence of external magnetic field the charged particles motion is unable to shield the displacement current. This is because  
the charges cannot simply get accelerated along the  electric field, when an ambient magnetic field perpendicular to the electric field is present. The presence of 
magnetic field curves the trajectories of the 
charge particles and makes them drift along the 
$\vec{E} \times\vec{B}_0$ direction. Where $\vec{B}_0$ is the external magnetic field. Thus, even in the low frequency 
domain the   displacement current does not get screened. 
  This is responsible for   the  existence and propagation of low frequency electromagnetic 
 Alfven and whistler wave   in magnetized plasmas \cite{freidberg87,kingsep}.  The Alfven wave corresponds to frequencies lower than the ion gyrofrequencies making both ion and electron 
 species as magnetized. The Whistler wave on the other hand has frequencies between ion and electron gyrofrequency, and in this case only the electrons 
 species is magnetized. The possibility of  shielding by unmagnetized ions  in this case is possible   provided the frequency of the EM wave is lower than the ion 
 plasma frequency. 
 \par\vspace{\baselineskip}

 The application of external magnetic field in  plasmas is often not feasible. In this manuscript, therefore, we look for alternative mechanism 
 aided  by the nonlinearity of the plasma medium to promote the propagation of slow  EM disturbance  through the plasma medium.  We restrict here to 
 that  regime of time variations in which the response from the heavy ion species  can be ignored. They merely provide a static neutralizing background charge  
 for the plasma medium. In other words we confine our discussions here to the time scale regime of 
 Electron Magnetohydrodynamics (EMHD) model \cite{kingsep,das_emhd,amita08,amita2001,pos,biskamp}.  
The model also neglects the displacement current under the assumption that the 
plasma is overdense ($\omega_p/\omega > 1$ and $\omega_p^2/\omega \omega_c > 1$ ) and the conduction current density far exceeds the displacement current. 
Under these conditions the electron  motion is incompressible and its vorticity equations can be cast  in terms of a set of equations for the magnetic field  $\vec{B}$.  
\par\vspace{\baselineskip}

\begin{figure}[!hbt]
 \centering
 \includegraphics[width=15cm,height=10cm]{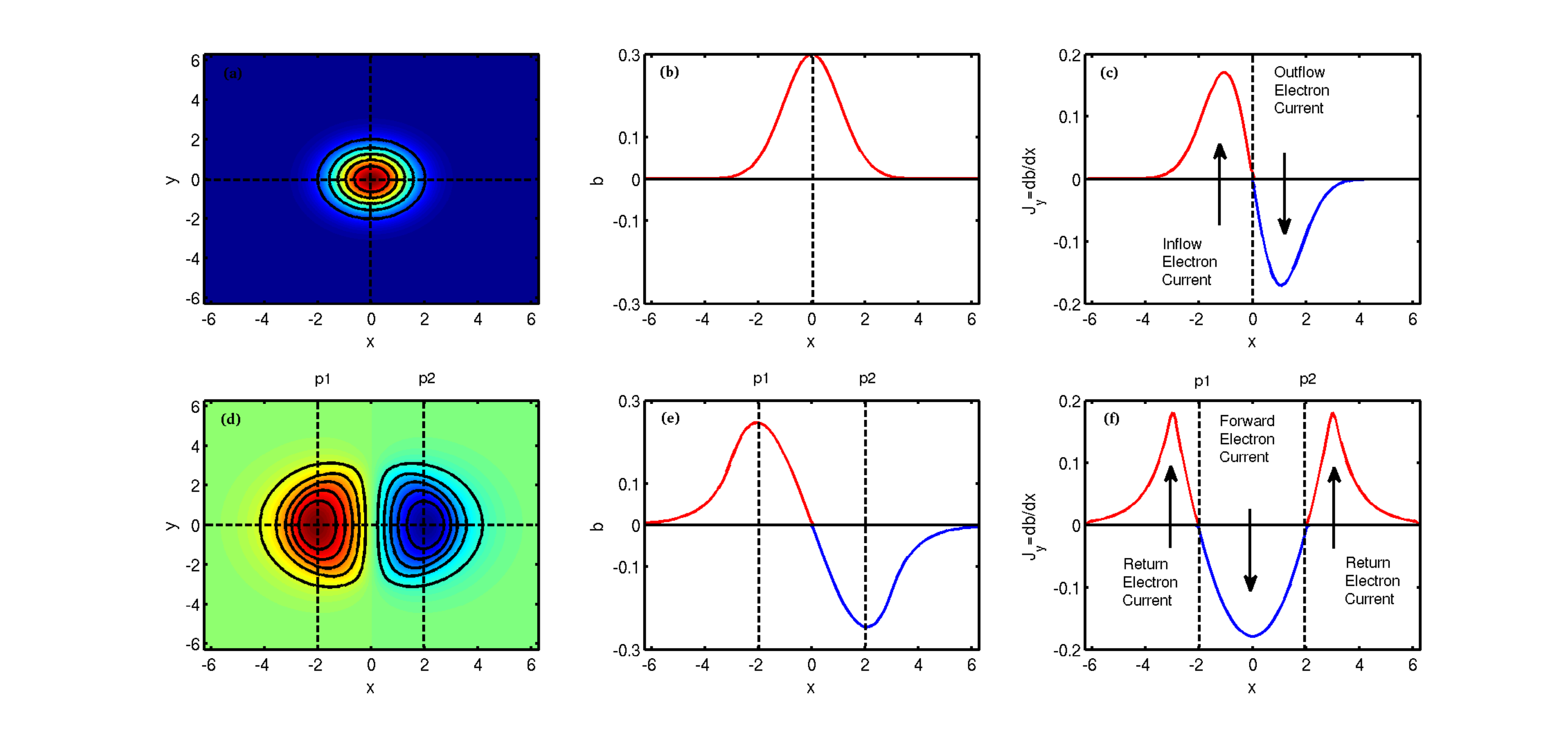}

  \caption{Schematic diagram of monopole and dipole magnetic field structures. The subplots (a) and (d) show the 
magnetic field contours and the color plot for monopoles and dipoles respectively. The subplot (b) and (c) 
show the line out profiles for the magnetic field $b$ and the electron current flow respectively at $y = 0$  as a function of $x$ for monopole magnetic field structures. The subplot (e) and (f) 
show the line out profiles for the magnetic field $b$ and the electron current flow respectively at $y = 0$  as a function of $x$ for dipole magnetic field structures.}
  \label{fig1}
 \end{figure}

\begin{figure}[!hbt]
 \centering
 \includegraphics[width=15cm,height=5cm]{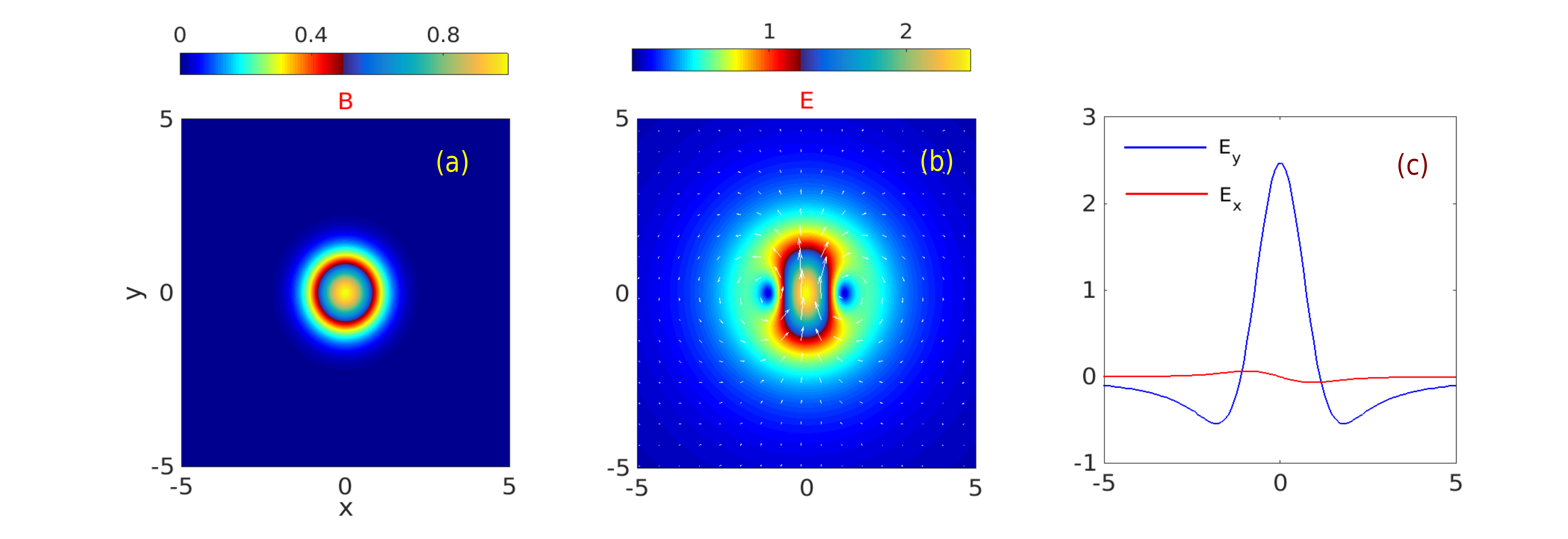}
  \caption{Schematic diagram of (a) monopole magnetic field structure (b) corresponding electric field structure, and (c) line out profiles of $x$ and $y$ component of electric field.}
  \label{fig2}
 \end{figure}

\begin{figure}[!hbt]
 \centering
 \includegraphics[width=15cm,height=5cm]{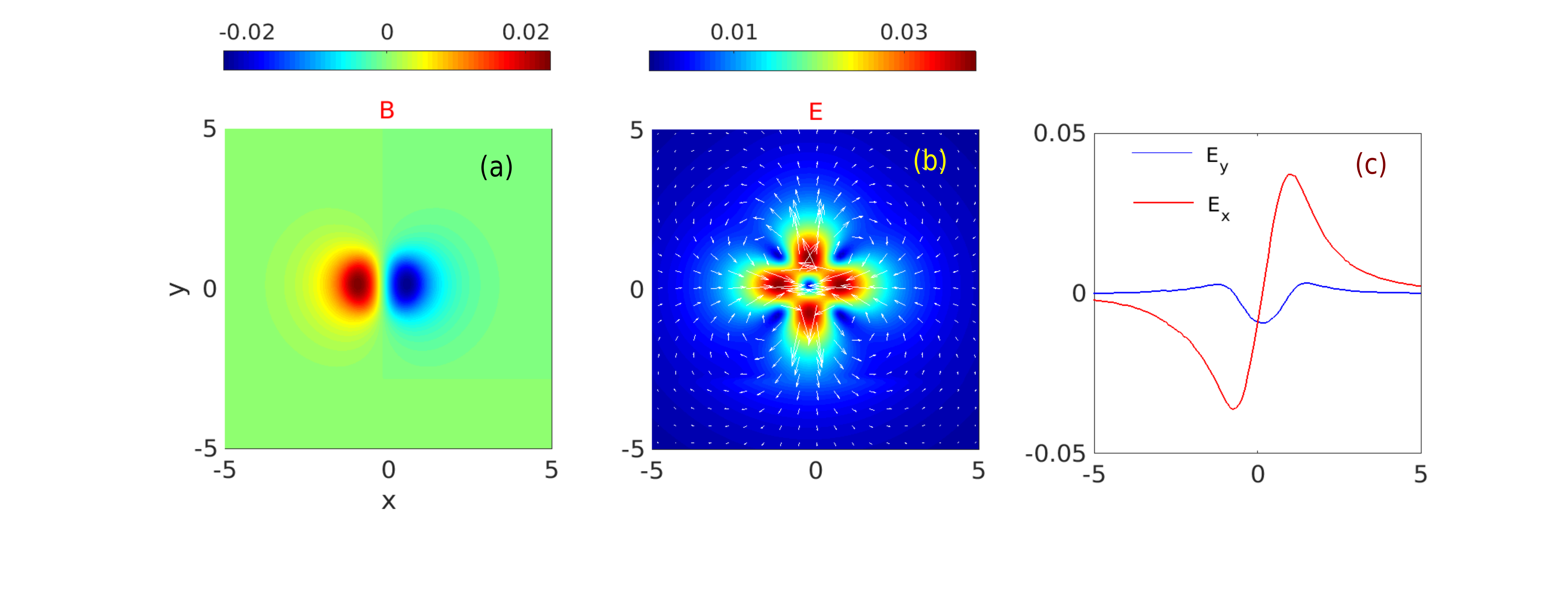}
  \caption{Schematic diagram of (a) dipole magnetic field structure (b) corresponding electric field structure, and (c) line out profiles of $x$ and $y$ component of electric field.}
  \label{fig3}
 \end{figure}

These equations posses exact nonlinear solutions describing current pulse structures and its associated magnetic field in a plasma \cite{isichenko}. Two kinds of structure are of special interest. 
  They are two dimensional structures having magnetic fields with monopolar and dipolar symmetries depicted in Fig.(\ref{fig1}).  
  The top   three 
subplots (a), (b) and (c) show the   contour plot of the associated magnetic field in 2-D plane, the profile of magnetic field and the electron flow at the mid $y = 0$ section of the 
structure respectively for the monopolar electron current pulse. 
These are radially symmetric rotating electron current flow patterns which are non - propagating in a homogeneous plasma. 
The subplots (d), (e) and (f) of Fig.(\ref{fig1}) corresponds to the same features for the dipolar solutions which 
  move with uniform axial speed  $U$  in a homogeneous plasma. The speed $U$ typically increases with the maximum   amplitude  of $\mid b \mid$ shown by the 
peak value in subplot(e) and it also increases with the increasing proximity of the two lobes.  This dipolar solution 
 can thus be considered as a  model for the 
finite propagating electron current pulse in the plasma for our studies.  It may also be viewed as a propagating magnetic field disturbance in an overdense plasma 
which is screened in the radial and axial directions by typical scale length of the order of electron skin depth. 
For these dipolar  structures  the central region (subplot(f)) shows a 
 forward (along the propagation direction) current 
 flow  which bifurcates and returns 
along both  sides as a return current.  As stated earlier this  entire structure propagates with a constant speed $U$ in the  plasma.  
  The Electric field lines in the plane associated with these monopolar and dipolar structures have also been shown in Fig. Fig.(\ref{fig2}) and Fig.(\ref{fig3}) respectively. 
  In the first subplot of these figures the line out of the two component of electric fields as a function of $..$ coordinate has been shown. The second on the other hand shown the electric field lines in the 2-D plane. 
\par\vspace{\baselineskip}

While the monopolar structures remain static in the medium, the dipolar structures are of great interest. 
  To a stationary observer this propagating dipole would appear as a time dependent Electromagnetic pulse having a Doppler shifted frequency 
 of $kU$, where $k = 2\pi/d$,  $d$ being the scale of the  dipolar structure. The scale length of such dipolar structures are typically of the order of 
 electron skin depth $d \sim c/\omega_{pe}$. Thus the associated frequency $kU = \omega_{pe}U/c << \omega_{pe}$. 
 These dipoles thus may be viewed as slowly moving exact nonlinear EM disturbances. EMHD simulations have shown their steady and stable 
 propagation \cite{das_ppcf}. 
 \par\vspace{\baselineskip}

 A natural question that arise in this context is whether such slow EM structures 
can be made to propagate inside an even higher density plasma region. In a recent study it was shown that it is 
indeed possible \cite{sharad08} for these structures to penetrate and propagate in higher density regions of plasmas. 
A generalization of the   EMHD 
model (G-EMHD) which incorporates the spatially inhomogeneous plasma densities  \cite{sharad08,sharad09,sharad10}  
was used to carry out the simulations. A  simplified 2-D case when the  symmetry axis is along $\hat{z}$ and the electron flow is  
confined to  the 2-D $x-y$  plane, 
the G-EMHD model can be cast  in terms of a single scalar field $b$ representing  the  sole 
component of magnetic field along the symmetry direction due to the 2-D electron current pulse.  
The evolution equation for $b$ can then be written as \cite{sharad08,sharad09,sharad10}: 
\begin{equation}
 \fpar{}{t}\left\{b - \nabla \cdot\left(\frac{\nabla b}{n} \right) \right\} + 
 \hat{z} \times \nabla b \cdot \nabla 
\left[\frac{1}{n}\left\{b - \nabla \cdot\left(\frac{\nabla b}{n} \right) \right\}\right]  = 0 
\label{2db}
\end{equation}
The equation has been written in normalized variables. The   
 magnetic field  and the electron density have been normalized by some typical 
values $B_{00}$ and $n_{00}$ respectively. Time is normalized by the corresponding electron gyroperiod 
$\omega_{ce0}^{-1} = 1/(e B_{00}/mc)$, length by the electron skin depth $d_{e0} = c /\omega_{pe0}$ 
(where $\omega_{pe0} = 4 \pi n_{00} e^2 /m$).  
When the density $n$ is uniform, the Eq.(\ref{2db}) reduces to a simpler form of EMHD model. 
\par\vspace{\baselineskip}

In the presence of a density inhomogeneity the magnetic field patterns associated with these current pulses
 acquire  an additional drift 
velocity $\vec{v}_d = \vec{B} \times \nabla n/n^2$ which is clearly transverse to the density 
gradient as well as the direction of magnetic field $\vec{B} = b \hat{z}$.  The monopolar current pulses 
which are otherwise non - propagating in a plasma 
can   thus be made 
to move along the contours of constant plasma density. By  altering the sign of $\nabla n$ the 
direction of propagation of the monopolar pulse can be reversed.  However, the monopolar current profiles 
are pretty restricted in terms of their maneuverability.    
They cannot move   
 across the density gradient.  The dipolar form of current profiles, which mimic the combination of forward and return 
 currents allows for a far superior maneuverability. 
 The 
 drift associated with the density gradient in conjunction with the intrinsic axial drift 
of the dipoles allows easy  penetration in higher density regions. 
This has been clearly illustrated in the simulations 
 carried out by Sharad et al., \cite{sharad08,boris} using Eq.(\ref{2db}).  It has  been shown that the EMHD dipole solution travels through the region of 
increasing density and can easily enter a high density region where it continues its  steady propagation. 
The structure scale length is observed to typically adjust to the skin 
depth of the high density region.  
\par\vspace{\baselineskip}

An interesting aspect of these structures are that the energy content of the pulse is equipartitioned between the field 
and the electron kinetic energy.  In the context of linear EM waves propagating in the plasma  it is typically partitioned in the ratio of $\omega_{pe}^2/\omega^2$. 
Thus the propagating pulse also carries a significant fraction of energetic electrons along with it. Thus, the pulses provide for the propagation of not 
merely EM energy but energetic electrons also along with it. 
In the next section we discuss some other interesting and novel features of such pulses as they propagate in an inhomogeneous plasma medium. 
\section{Applications} 
\begin{figure}[!hbt]
 \centering
 \includegraphics[width=15cm,height=15cm]{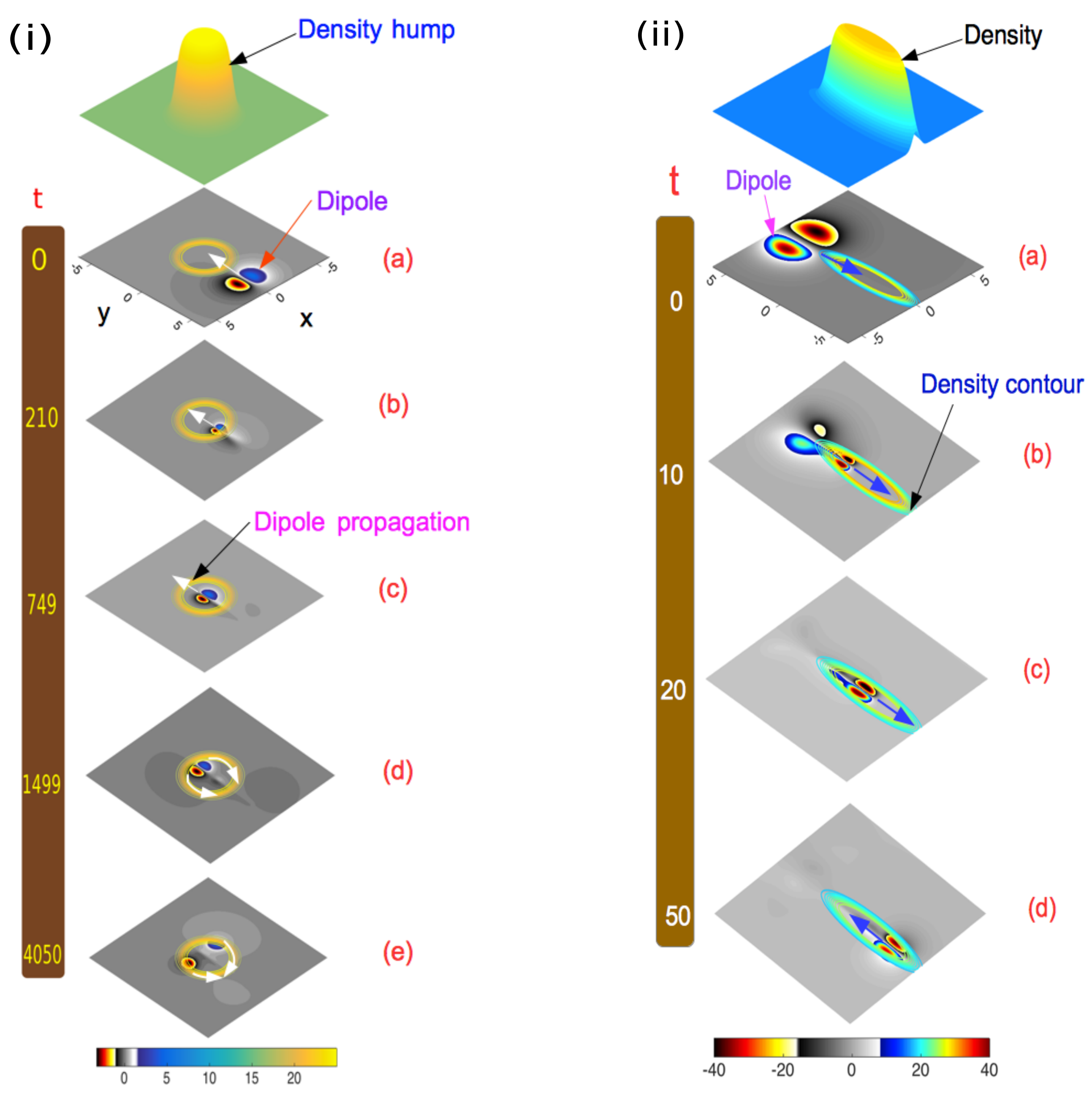}
  \caption{Propagation of magnetic dipole structure  (i) in a circular shaped density hump profile (ii) in a elliptic shaped density hump profile,  at different times.}
  \label{fig4}
 \end{figure}

We now discuss some other  interesting aspects of the propagation of these dipolar EM pulses through an appropriately tailored inhomogeneous electron density profile. 
It has  been shown in  simulations by Sharad et. al. \cite{sharad08,sharad09,sharad10} that once the pulse enters a high density region 
remains trapped inside it. Thus, an appropriate choice of plasma density profile can be utilized to trap the EM pulse inside it. 
In fact the high density plasma can act as a tweezer by confining the pulse inside it. This has been illustrated in Fig.(\ref{fig4}(i)), where we have chosen a high density circular hump of plasma density.
The electromagnetic pulse enters inside it and then gets trapped within the 
high density structure. By changing the density profile and making it narrower in one dimension the EM pulse gets collimated 
as has been shown in Fig.(\ref{fig4}(ii)). 
\par\vspace{\baselineskip}

This is  a very  
attractive proposition as  a simple choice of a narrow  high density plasma 
can suitably focus a divergent flow of EM disturbance as well as the electrons associated with the current in such pulses.  
In the context of many experiments  for the focused transport of electrons this attribute would be very desirable. 
\par\vspace{\baselineskip}
The observed  features can be easily interpreted in terms of an interplay of the two drifts associated 
with the dipolar current pulse. The drift associated with the density gradient brings the two lobes 
 with opposite polarity of the magnetic field together as  the dipole approaches the 
 high density region. This results in the collimation of the current pulse structure. 
 The collimated structure moves with greater axial speed and penetrates the high density region of the plasma.  
 Once inside the high density region the current pulse  propagates along  it to reach the other end 
 through the axial dipolar drift. 

\par
\begin{figure}[!hbt]
 \centering
 \includegraphics[width=15cm,height=15cm]{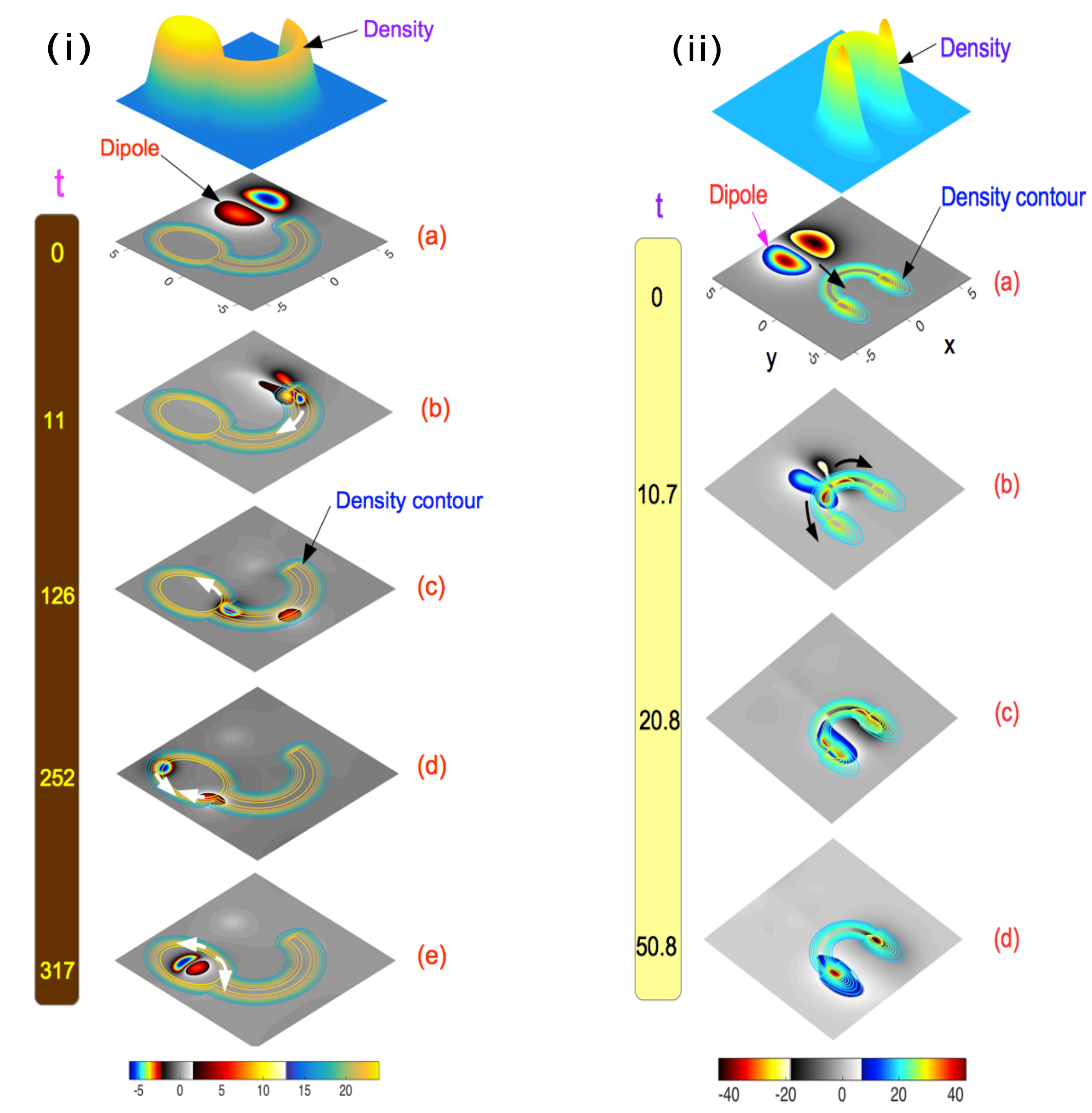}
  \caption{Propagation of magnetic dipole structure in a curved shaped density hump profile for (i)  guiding the pulse (ii) bifurcation of the pulse, at different times.}
  \label{fig5}
 \end{figure}

\vspace{\baselineskip}
In the case of Fig.5 we had shown a straight propagation of the current pulse towards the other end of the narrow 
high density plasma region. We have also carried out  simulations for a 
 curved  high density region as indicated by the  contours in Fig.(\ref{fig5}(i)). 
 The current pulse trajectory in this case gets suitably altered to get guided along the path defined by the high density narrow plasma  region. 
 We  thus see  that  by choosing appropriately tailored and  
 different forms of the plasma density 
 the dipole current pulse can be 
guided  and sent to a desired destination.  In  Fig.(\ref{fig5}(i)) in fact one has been able to reverse the propagation direction of the original pulse.  
\par\vspace{\baselineskip}
The question as to how narrow should the density inhomogeneity be for the guiding to occur is settled from the considerations of 
the fact that the electron current pulse should be able to feel the inhomogeneity of the density region. 
If the high density region to which the dipole pulse enters is considerably broader than the structure size of the pulse, the pulse will not feel the 
edges and will not experience the density inhomogeneity related drift.  The pulse size on the other hand gets suitably adjusted to the local 
skin depth of the associated plasma. Thus the width of  the path along which the pulse is guided  should be smaller compared to the typical 
skin depth size. 
\par\vspace{\baselineskip}
The observed collimated propagation of current pulse through G-EMHD simulations along the direction defined 
by the  narrow high density plasma region forms the underlying physical basis of some  experimental observations.  
For instance, in  an experiment reported by Kodama {\it et al.} \cite{kodama_wire}  the 
 fast electrons get generated by impinging ultra intense laser 
pulse on a target in the shape of a gold cone. A  fine carbon wire was attached at the tip of the cone.  
It was shown in the experiment that the 
 electrons followed  the path defined by the direction of the solid carbon wire. 
When the wire was tilted with respect to the cone axis the electrons 
hit the imaging plate target  at an off axis location defined by the tilt of the wire. 
 The experiment can be understood on the basis of our mechanism. The wire gets ionized by the front of 
 the energetic electron pulse,  creating  a 
 narrow high density plasma region of the shape of the wire. 
The subsequent part of the electron 
 pulse then gets guided along this inhomogeneous plasma as proposed by us. 

\par\vspace{\baselineskip}
We now provide another possibility in connection with maneuvering the propagation of the current pulse path.
We show that one can  also bifurcate a current pulse which arise from the same source.  The 
 two parts can then be made to propagate and reach altogether different destinations. This has been shown 
by the  snapshots at various times for the current pulse structure through a density inhomogeneity of the kind 
shown by the thick black lines of Fig. (\ref{fig5}(ii)). The  lines  show the contours of the high density region. 
As the pulse enters the high density region, it  gets separated in two parts which then the two propagate 
along different directions.  
\section{Summary and discussion}
It is well known that low frequency (lower than plasma frequency) EM waves cannot propagate inside an unmagnetized plasma \cite{dendy90}. 
We show that the dipolar current solutions of the EMHD equations can in fact be treated as a  low frequency EM disturbances 
which can propagate inside a high density plasma aided by the inherent nonlinearity. 
These pulses can easily be pushed in the domain of even higher plasma densities. The energy content of these pulses are 
typically equipartitioned between the electron kinetic and field energies.  Thus the structure  can be an effective tool for the transport of 
energy in denser regions of the plasma. 
\par\vspace{\baselineskip}
A host of  frontline physics experiments today are associated  with 
 generation and/or the utilization of
 energetic electrons.  
One is either chasing the possibility and devising new schemes for  the generation of electrons with desirable  energy range  (electron acceleration ) 
 or is interesting in utilizing them for a variety of   purposes. For instance, the energetic electrons are used  for the purpose of 
localized heating in hot dense plasma targets as in the  fast ignition experiments \cite{fi_revs,vehn_ppcf05,campbell,honorubia,honda},  or as a diagnostic tool for hot dense plasma studies etc. 
For such diverse applications one often encounters the question of keeping the electron pulse collimated and also its  controlled guided  propagation  along a desired path.  
This question has been addressed by a number of authors recently. For instance,  techniques which rely on 
structured target designs of different materials with varying resistivity has been used for the purpose of electron pulse guiding \cite{robinson_pop}. 
The strong magnetic field generated at the interface 
of materials having different resistivities is important   for  guiding  the current pulse in this case. 
The guiding due to such structured targets  has also  been experimentally 
verified in the work by Kar {\it et al.} \cite{kar_prl}. The preparation of structured targets, however,  may  
not be convenient and/or feasible for use  in all kinds  of  experiments.
The recent work by Kodama {\it et al.} \cite{kodama_wire} reported in 
Nature provides a simpler scheme wherein they are able to show that the electron pulse gets guided 
along the direction of solid wire placed in their path.  
The effects shown here, however, have been for low non relativistic electrons and need to be generalized 
for the relativistic case for it to be suitably applied for the applications mentioned above. such a generalization 
will be carried out and presented in a future publication. Here, we confine ourselves to some interesting physical insights 
on the nonlinear EM pulse propagation in plasma s aided by the electron flow in the medium. 
\par\vspace{\baselineskip}

Just as plasma photonics promises  a novel 
technique for guiding photons, we show here that the plasma medium with a tailored density profile offers  a simple and novel  scheme for guided electron transport. 
The   physics of electron current pulse propagation through plasmas is analyzed and then with the help of 
numerical simulations it is shown that the introduction of an appropriate density inhomogeneity leads to the guiding of electron current pulses.  
We  specifically show here that 
 electron current pulse can be collimated and guided along a desired path 
by a suitable choice of plasma density 
profile. We also demonstrate that a single electron current pulse can be  bifurcated  and send to 
distinct locations  in a plasma for dual usage  similar to the splitting of a  photon beam.  
The proposed technique is practically viable as the  appropriate plasma inhomogeneity in question can easily be created by the ionizing a solid target of appropriate 
density profile.  In fact in the 
   experiments by Kodama {\it et al.} \cite{kodama_wire} the energetic electron itself 
generates the requisite plasma inhomogeneity profile by ionizing the wire through which it propagates. 

\noindent

\newpage


\end{document}